\documentclass[letter,times,twocolumn,tighten]{aastex62}
\hypersetup{linkcolor=red,citecolor=blue,filecolor=cyan,urlcolor=magenta}
\usepackage{amsmath,amstext}
\usepackage[T1]{fontenc}
\usepackage{apjfonts}
\usepackage[figure,figure*]{hypcap}
\usepackage{commath}
\usepackage{booktabs}

\definecolor{RoyalBlue}{rgb}{0.25,0.41,0.88}
\definecolor{Red}{rgb}{0.61,0.12, 0.14}
\definecolor{GoldenRod}{rgb}{0.8,0.40, 0.}

\newcommand{\revision}[1]{#1}

\shorttitle{Resolving FU~Ori's twin disks}
\shortauthors{Perez et al.}

\begin{document}

\title{RESOLVING THE FU ORI SYSTEM WITH ALMA: INTERACTING TWIN DISKS?}

\correspondingauthor{Sebasti\'an P\'erez}
\email{sebastian.astrophysics@gmail.com}
\author[0000-0003-2953-755X]{Sebasti\'an P\'erez}

\affil{Universidad de Santiago de Chile, Av. Libertador Bernardo O'Higgins 3363, Estaci\'on Central, Santiago}
\affil{Departamento de Astronom\'ia, Universidad de Chile, Casilla 36-D, Santiago}

\author[0000-0001-5073-2849]{Antonio Hales}
\affiliation{National Radio Astronomy Observatory, 520 Edgemont Road, Charlottesville, VA 22903-2475.}
\affiliation{Joint ALMA Observatory, Alonso de C\'ordova 3107, Vitacura 763-0355, Santiago}

\author[0000-0003-2300-2626]{Hauyu Baobab Liu}
\affiliation{Academia Sinica Institute of Astronomy and Astrophysics, P.O. Box 23-141, Taipei 10617, Taiwan}

\author[0000-0003-3616-6822]{Zhaohuan Zhu}
\affiliation{Department of Physics and Astronomy, University of Nevada, Las Vegas, 4505 S. Maryland Pkwy, Las Vegas, NV 89154, USA}

\author[0000-0002-0433-9840]{Simon Casassus}
\affil{Departamento de Astronom\'ia, Universidad de Chile, Casilla 36-D, Santiago}

\author[0000-0001-5058-695X]{Jonathan Williams}
\affiliation{Institute for Astronomy, University of Hawaii, Honolulu, HI 96822, USA}

\author[0000-0002-5903-8316]{Alice Zurlo}
\affiliation{N\'ucleo de Astronom\'ia, Facultad de Ingenier\'ia y Ciencias, Universidad Diego Portales, Av. Ejercito 441, Santiago, Chile}
\affiliation{Escuela de Ingenier\'ia Industrial, Facultad de Ingenier\'ia y Ciencias, Universidad Diego Portales, Av. Ejercito 441, Santiago, Chile}

\author[0000-0003-3713-8073]{Nicol\'as Cuello}
\affiliation{Instituto de Astrof\'isica, Pontificia Universidad Cat\'olica de Chile, Vicu\~na Mackenna 4860, 7820436 Macul, Santiago}
\affiliation{N\'ucleo Milenio de Formaci\'on Planetaria (NPF), Chile}

\author[0000-0002-2828-1153]{Lucas Cieza}
\affiliation{N\'ucleo de Astronom\'ia, Facultad de Ingenier\'ia y Ciencias, Universidad Diego Portales, Av. Ejercito 441, Santiago, Chile}

\author[0000-0002-7939-377X]{David Principe} 
\affiliation{Massachusetts Institute of Technology, Kavli Institute for Astrophysics, Cambridge, MA, USA}

\begin{abstract}
  \revision{FU~Orionis objects are low-mass pre-main sequence stars
    characterized by dramatic outbursts of several magnitudes in
    brightness. These outbursts are linked to episodic accretion
    events in which stars gain a significant portion of their
    mass. The physical processes behind these accretion events are not
    yet well understood. The archetypical FU~Ori system, FU~Orionis,
    is composed of two young stars with detected gas and dust
    emission. The continuum emitting regions have not been resolved
    until now. Here, we present 1.3~mm observations of the FU~Ori
    binary system with ALMA. The disks are resolved at 40~mas
    resolution. Radiative transfer modeling shows that the emission
    from FU~Ori north (primary) is consistent with a dust disk with a
    characteristic radius of $\sim$11~au. The ratio between major and
    minor axes shows that the inclination of the disk is
    $\sim$37$^{\circ}$. FU~Ori south is consistent with a dust disk of
    similar inclination and size. Assuming the binary orbit shares the
    same inclination angle as the disks, the deprojected distance
    between north and south components is 0\farcs6,
    i.e. $\sim$250~au. Maps of $^{12}$CO emission show a complex
    kinematic environment with signatures disk rotation at the
    location of the northern component, and also (to a lesser extent)
    for FU Ori south. The revised disk geometry allows us to update FU
    Ori accretion models (Zhu et al.), yielding a stellar mass and
    mass accretion rate of FU~Ori~north of 0.6~M$_{\odot}$ and
    3.8$\times10^{-5}$ M$_{\odot}$ yr$^{-1}$, respectively.}
\end{abstract}

    

\section{Introduction} \label{sec:intro}

Protostars build a significant fraction of their final mass through an
accretion disk fed by a surrounding envelope. As this envelope
dissipates, accretion rates tend to drop below 10$^{-7}$ or even
10$^{-8}$~M$_\odot$\,yr$^{-1}$. While most young stellar objects
(YSOs) have luminosities that are significantly lower than expected
from steady protostellar disk accretion \citep[the so-called {\em
    luminosity problem;}][]{Ken1990, Eva2009}, some exhibit episodes
of high activity, increasing their optical brightness by several
orders of magnitude on time-scales as short as one year
\citep[see][for a review]{Audard2014}.  Accretion at this stage can
reach up to 10$^{-4}$~M$_\odot$\,yr$^{-1}$. A possible solution to the
so-called luminosity problem is that most protostars also go through
an `FUor' (named after the class prototype FU Orionis) phase early in
their life time. This would imply that a significant fraction of the
mass of a young protostar is accreted in an episodic way, with
prolonged periods of low accretion \citep{Zhu2009, Vor2010, Bae2014}.

An Early Science campaign to characterize FUor sources with the
Atacama Large Millimeter/sub-millimeter Array (ALMA) shows that they
are compact in radio continuum \citep{Cieza2018}. Their disk emission
have smaller characteristic radii for a given disk mass than T~Tauri
disks and have disk sizes and masses that are more similar to those of
Class I than to Class II sources
\citep[e.g.,][]{sheehan2017,tobin2018}. FUors appear surrounded by
prominent outflows of molecular gas with hourglass-like morphologies
\citep[e.g., V2775~Ori,][]{Zurlo2017}. These outflows can sometimes
display wide opening angles \citep[e.g.,
  HBC~494,][]{Ruiz-Rodriguez2017a, Ruiz-Rodriguez2017b}, and even show
notable misalignments with respect to the orientation of the dust disk
\citep[e.g., V1647~Ori,][]{Principe2018}.

Episodic accretion and its implications for star and planet formation
are not well understood. Several physical processes have been proposed
to explain such dramatic accretion events. The most favored mechanisms
include disk fragmentation and the subsequent inward migration of the
fragments \citep{Vor2010}, gravitational instability (GI) and
magneto-rotational instabilities \citep{Armitage2001, Zhu2009,
  Zhu2010, Martin2011, Bae2014}, amongst others. These models predict
the presence of distinct features (such as clumps and spiral arms)
that should be detectable with current observational capabilities,
although these have not been found on high resolution images of FUor
disks \citep[e.g. V883~Ori,][]{cieza2016}.

Another possibility is that, since these young disks dwell in crowded
star-forming regions, their outbursts may be connected to, or driven
by, binary interactions \citep{Bonnell1992, Reipurth2004}, or
star-disk flybies \citep[e.g.,][]{Pfalzner2008, Cuello2019} as stars
are common and disks have large cross-sections, and possibly even
disk-disk interactions \citep[e.g.,][]{Mun2015}. A few FUor-like
protostars have now been identified as having binary components
\citep[e.g., L1551~IRS5, RNO~1B/C, AR~6A/B,][and references
  therein]{Pue2012}, feeding the discussion on whether the rapid
accretion could be caused by close-companion
interactions. \revision{Perturbers such as planets have also been
  proposed as an explanation for the quick rise in the outburst
  lightcurves of FU Ori and V1057 Cyg \citep{Clarke2005}.}
 
ALMA Early Science observations of the FU~Ori binary system show that
each binary component is surrounded by compact continuum emission
(unresolved at $\sim$0\farcs6 resolution) and widespread molecular
emission \citep{Hales2015}. Using the Karl G. Jansky Very Large Array
(JVLA) in its most extended configuration, \citet{Liu2017} separate
the two binary disks, but the individual disks remain unresolved at
29~GHz with a 0\farcs07 resolving beam. Combining these observations
with lower (8-10~GHz) and higher frequency data, \citet{Liu2017} show
that the 8 to 346~GHz SEDs of FU Ori North (or simply ``FU~Ori'', the
primary) and FU Ori South (secondary) cannot be reproduced by constant
spectral indices, suggesting a multi-component circumstellar
environment (bearing in mind that these sources are time
variable). \citet{Liu2017} also show that the observations can be
explained by a circumstellar environment consisting of free-free
emission from ionized gas, and thermal emission from two dust
components, a compact optically thick disk and an optically thin
region. On the other hand, \citet{Zhu2007} show that the optical-IR
spectrum of FU~Ori requires a hot disk with mass accretion of
$2\times10^{-4}$~M$_\odot$\,yr$^{-1}$.

In this work, we present 40~mas \revision{resolution} ALMA
observations of the FU~Ori system which resolve the continuum emitting
regions around each stellar component, providing the first direct
measure of their sizes at millimeter wavelengths
(Section~\ref{sec:modeling}). An analysis of the local kinematics
around the FU~Ori components using molecular line observations is
presented in Section~\ref{sec:kine}. Implications are discussed in
Section~\ref{sec:results}, while Section~\ref{sec:conclusions}
presents our concluding remarks.  Throughout this work we use a
distance to FU~Ori of 416$\pm$9~pc \citep[][DR2]{GAIA2018}, consistent
with the distance to the Orion Nebula Cluster from parallax of
non-thermal radio emission \citep{Menten2007}.

\section{Observations and data reduction} \label{sec:observations}

FU~Ori was observed during ALMA Cycle 4 on September 6$^{\rm th}$ 2017
using ALMA's 12m array in extended (C40-8) configuration, with 42
antennas with baselines ranging from 41~m to 7.5~km. This
configuration yields an angular resolution of $\sim$0\farcs05
($\sim$21~au) and a Maximum Recoverable Scale of $\sim$1\farcs05
($\sim$437~au) at 1.3~mm. The median precipitable water vapor column
(PWV) was 0.35~mm, with clear sky conditions and an average phase RMS
of 28.6$^{\circ}$ after WVR correction.

The primary flux calibrator was the quasar J0423$-$0120, while
J0510+1800 was used as bandpass calibrator. The quasar J0532+0732,
located 3.5$^{\circ}$ from the target was observed as phase
calibrator, by alternating with the science target every 54 seconds to
calibrate the time-dependent variation of the complex gains. The total
time spent on-source was 39.8 minutes. A secondary phase calibrator
(J0551+0829), located 4.7$^{\circ}$ from the phase calibrator, was
observed regularly as a check source to assess the quality of the
phase transfer. The ALMA correlator was configured in Frequency
Division Mode (FDM). Two spectral windows with 1.875~GHz bandwidth
were set up for detecting the dust continuum, centered at 232.005~GHz
and 218.505~GHz, respectively. The $^{12}$CO(2--1), $^{13}$CO(2--1)
and C$^{18}$O(2--1) transitions of carbon monoxide were targeted by
configuring three spectral windows at rest frequencies of 230.538~GHz,
220.399~GHz and 219.560~GHz respectively. The spectral resolution for
the line observations was 122.070~kHz ($\sim$0.2~km~s$^{-1}$).  All
data were calibrated by the ALMA staff using the ALMA Science Pipeline
version r39732 in the CASA package \citep{2007ASPC..376..127M}.

\subsection{ALMA continuum imaging}
\label{sec:image}

\begin{figure*}
  \centering\includegraphics[width=0.52\textwidth]{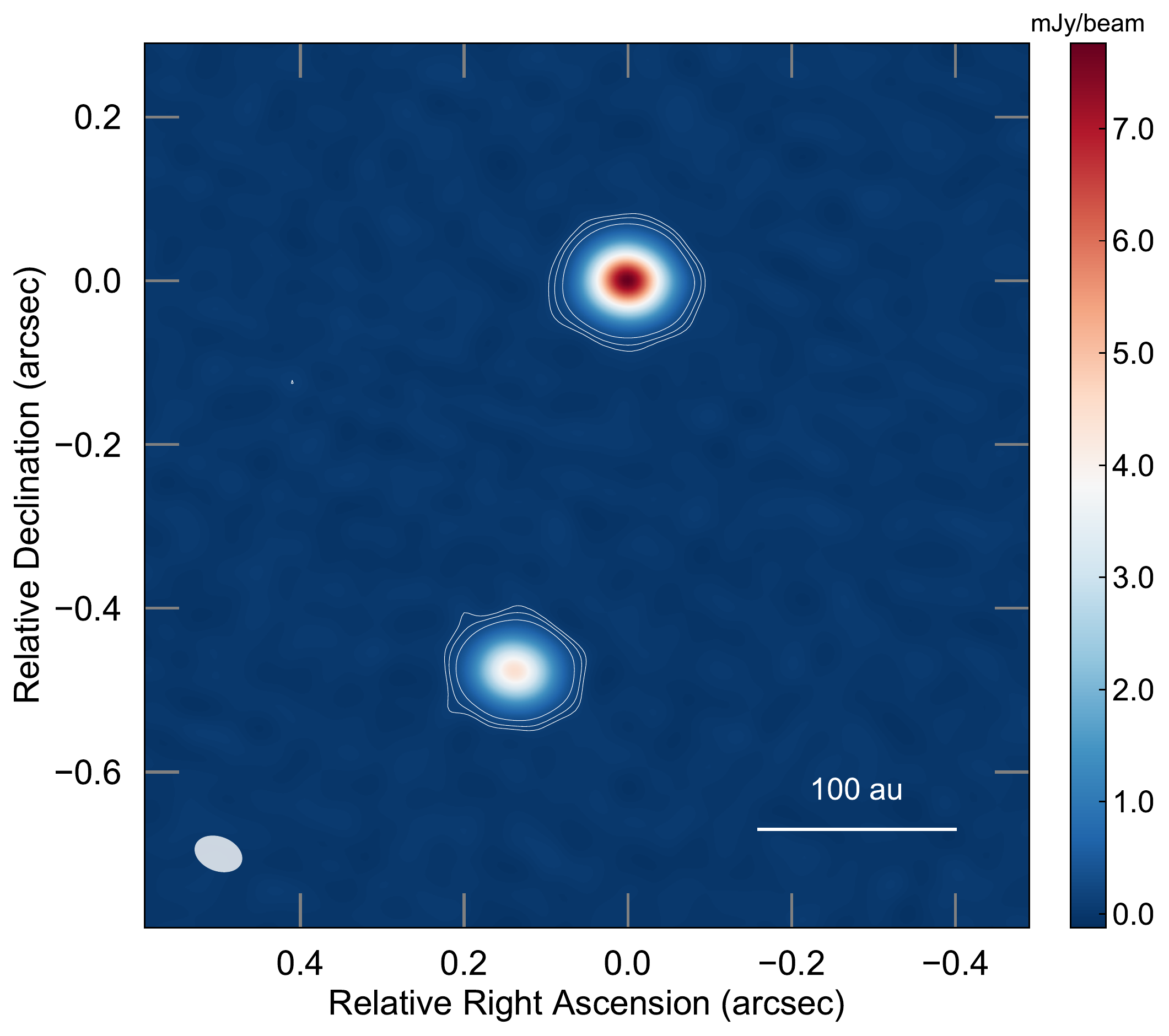}
  \hfill
  \includegraphics[width=0.46\textwidth]{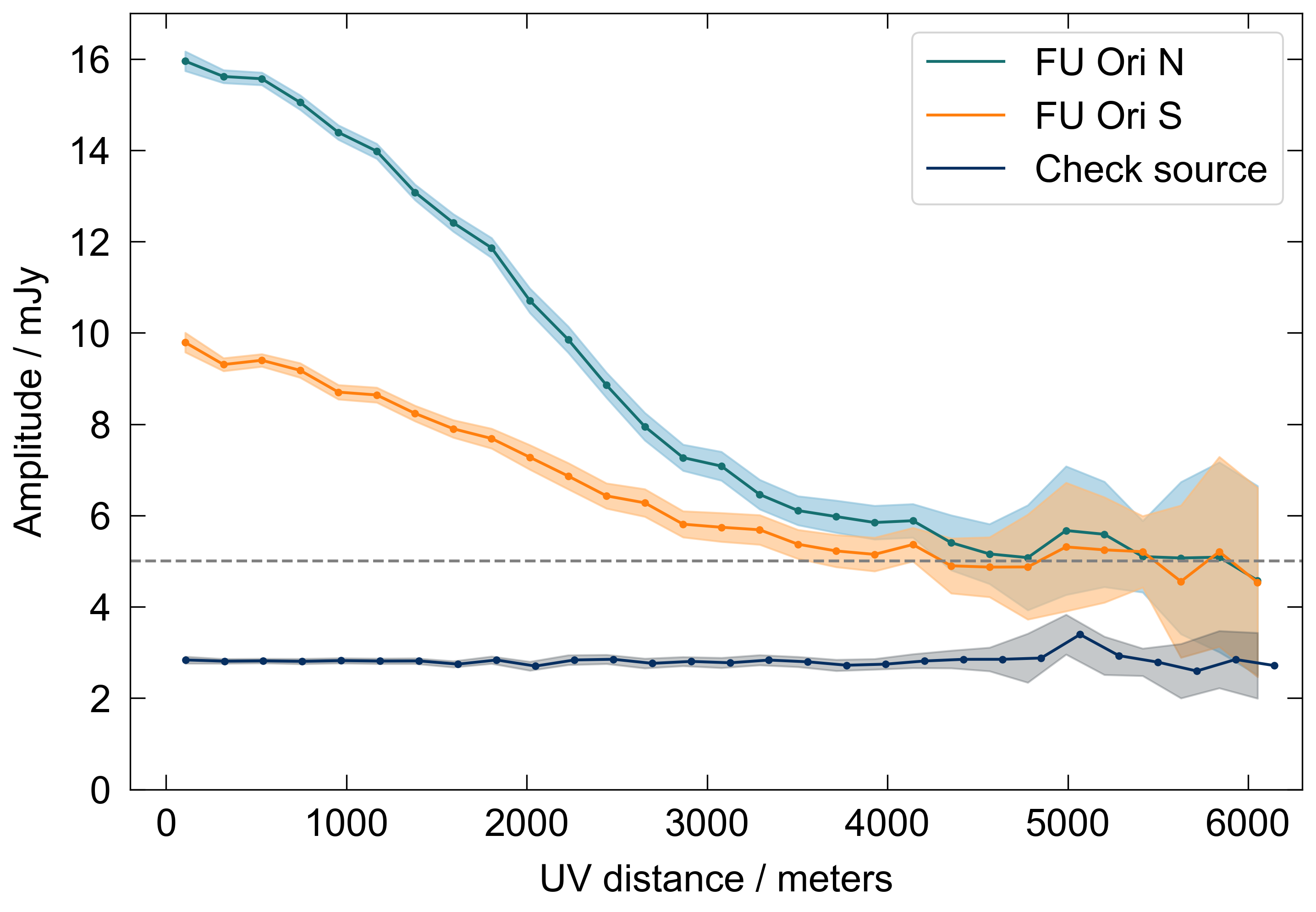}
  \label{fig:cont}
  \caption{{\em Left:} 1.3\,mm (225~GHz) continuum map of the FU~Ori
    binary system. The image resolution is 0\farcs06$\times$0\farcs042
    (shown in left bottom corner). The contours are 3, 5, and 10 times
    the image RMS noise of 30\,$\mu$Jy\,beam$^{-1}$. North is up, East
    is left. {\em Right:} Visibility amplitude as a function of
    baseline lengths (UV distances) for FU~Ori North (blue), FU~Ori
    South (orange) and check source J0551+0829, with their respective
    uncertainties. The visibilities of both FU Ori components have a
    profile that decreases from the peak maximum to a 5 mJy floor
    level (dotted line). The signal for J0551+0829 is constant with
    baseline length, as expected for an unresolved source.  }
\end{figure*}

Imaging of the 1.3~mm continuum emission was performed using the CLEAN
task in CASA with uniform weighting, resulting in a synthesized beam
size of 60$\times$42~mas and position angle of 68.8$^{\circ}$
(Figure~\ref{fig:cont}).  Self-calibration of the data was performed
to improve coherence. \revision{A single iteration of phase-only
  (i.e., no amplitude) calibration} resulted in a 1.7 factor
improvement in the resulting SNR (30~$\mu$Jy RMS) \revision{(compared
  to the SNR of a CLEAN image before self-calibration)}. Another
iteration of self-calibration was found to improve only marginally the
image quality.

As shown in Figure~\ref{fig:cont}, both components of the FU~Ori
binary are detected in continuum emission, separated by
$\sim$0\farcs5. FU~Ori north has a total \revision{flux density} of
14.30$\pm$1.5~mJy, while the southern component has a \revision{flux
  density} of 7.94$\pm$0.8~mJy (errors include the 10\% uncertainty in
the absolute flux calibration). The peak fluxes are 7.7$\pm$0.8 and
4.4$\pm$0.5~mJy~beam$^{-1}$ for FU Ori north and south,
respectively. The emitting regions around each component are resolved
at 60$\times$42 mas angular resolution. The deconvolved source sizes
(as FWHM) measured using the {\sc CASA} task {\tt imfit} are
50.4$\pm$0.5$\times$39.9$\pm0.6$~(mas) at position angle (PA)
133.6$^{\circ}$ $\pm$1.7$^{\circ}$ and
47.7$\pm0.8$$\times$38.4$\pm0.9$~(mas) at 137.7$^{\circ}$
$\pm$3.7$^{\circ}$ for FU~Ori~N and FU~Ori~S, respectively. This
corresponds to disk sizes of 7.5~au$\times$5.8~au and
7.2~au$\times$5.7~au, north and south, respectively.  The deconvolved
sizes translate to an inclination angle of
37.7$^{\circ}$$\pm$0.8$^{\circ}$ and 36.4$^{\circ}$$\pm$1.2$^{\circ}$,
for the north and south components, respectively. As will be shown in
Section~\ref{sec:kine}, the PA inferred from the continuum is in good
agreement with rotation seen in the kinematics of FU~Ori~N.

 
As visibilities are spatially-integrated quantities, one of the
components needs to be subtracted to inspect the visibilities of the
other. This is achieved by subtracting the CLEAN model of one of the
components from the visibility data. There are no indications of
substructures in either disk, or of emission in between the stellar
components in continuum. The observed visibilities of each component
(right panels in Fig.~\ref{fig:cont}) show the decreasing profile
characteristic of extended sources \citep[e.g.][]{Wil2009}.

By imaging each continuum spectral window separately (218 and 232
GHz), we derive in-band spectral indices (using the full bandwidth of
each spectral window, i.e., 1.875~GHz) of 2.1 for both FU~Ori and
FU~Ori~S. This is consistent with the spectral indexes derived in
recent SED fitting \citep{Liu2019}. The ALMA data confirm both disks
have almost identical spectral indices, consistent with the emission
being optically thick.

\subsection{ALMA $^{12}{\rm CO}$ imaging}

To construct channel maps of CO emission, a CLEAN procedure was
performed on the continuum-subtracted data. We used a Briggs weighting
scheme with a robust parameter of 1.0. This weighting yields the best
results in terms of achieving good signal-to-noise without
compromising on resolution. Channel maps were produced with a spectral
resolution equivalent to 1\,km\,s$^{-1}$. These broad channels are
needed to pinpoint fast Keplerian kinematics as opposed to slow
outflows.  Each channel map has an RMS noise of 1.2 mJy\,beam$^{-1}$,
for a CLEAN beam of 0\farcs1$\times$0\farcs09. \revision{The 1$\sigma$
  noise level is 2 mJy\,beam$^{-1}$ if systematics (large scale
  fringes in the central channels) are also included.} As shown in
\citet{Hales2015}, the kinematics is complex and it is heavily
influenced by large scale cloud emission and absorption, and also
possibly a slow outflow. A full analysis of the global kinematics,
i.e., how outflowing and cloud material connect to the binary FU~Ori
kinematics, will be presented in a future publication (Hales et
al. {\em in prep}). Here we focus on probing the kinematics of the gas
in the vicinity of the FU~Ori stars.

\section{Continuum modeling}
\label{sec:modeling}

\subsection{Radiative transfer}
\label{sec:rt}

\begin{figure*}
  \hfill\includegraphics[width=0.6\textwidth]{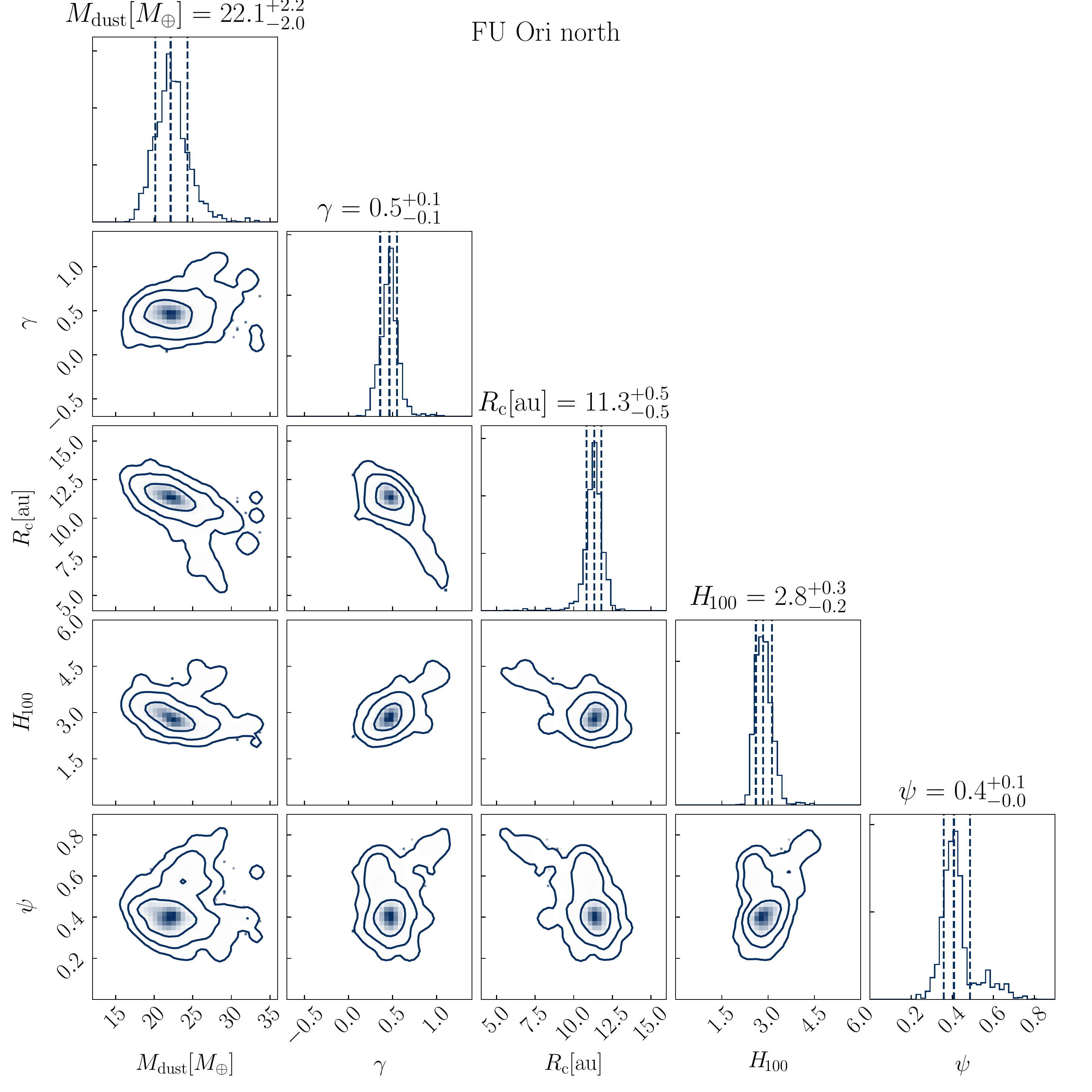}\\
  \includegraphics[width=0.6\textwidth]{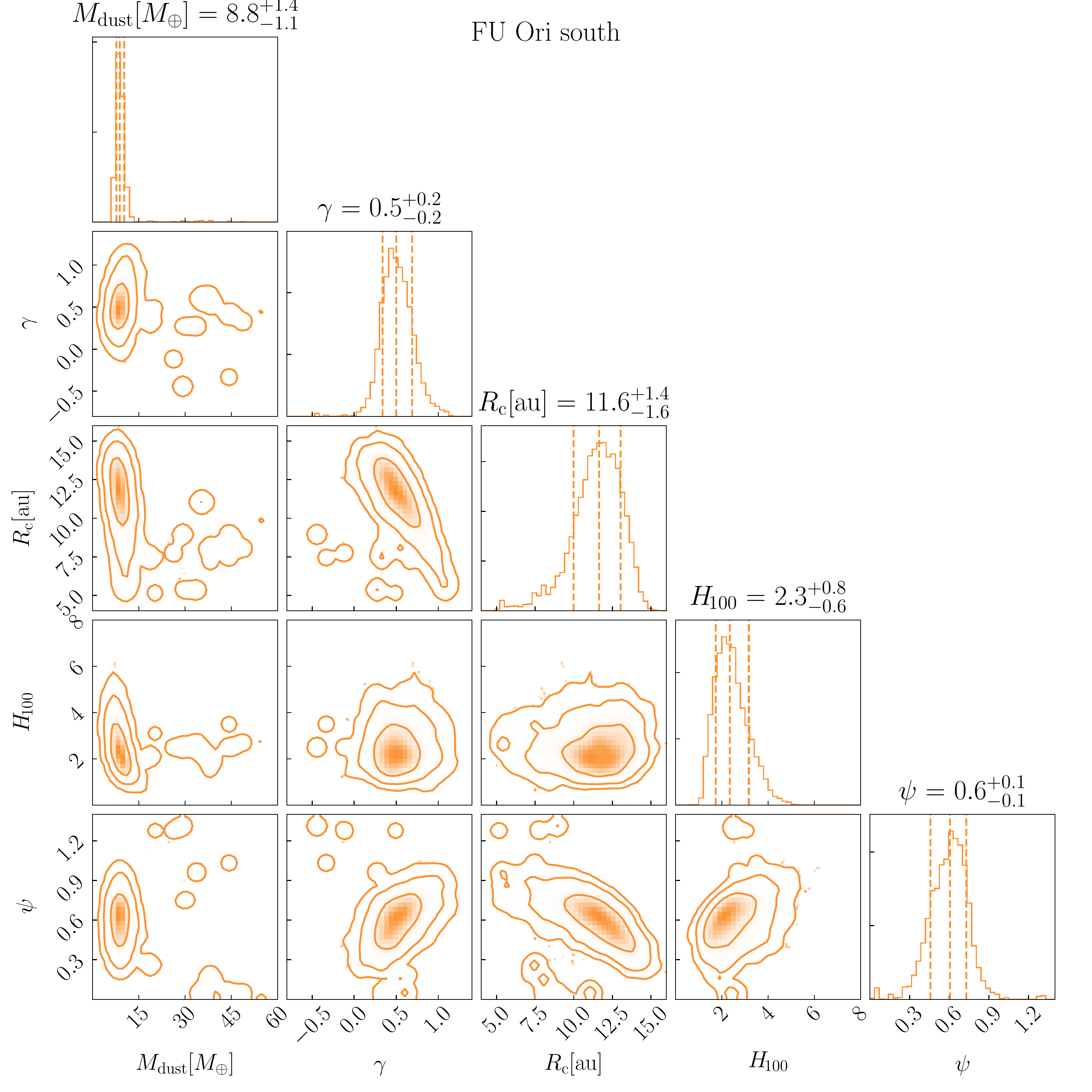}
  \caption{Posterior distributions of disk parameters $M_{\rm dust}$,
    $\gamma$, $R_c$, $H_{100}$, and flaring index $\Psi$, including
    their marginalised distributions for FU~Ori~north (top right,
    blue) and south (bottom left, orange). The vertical dashed lines
    represent the 16th, 50th and 84th percentiles. Contours correspond
    to 68 per cent, 95 per cent and 99.7 per cent confidence
    regions. Units are specified on the panel titles. The plots were
    generated using the {\sc python} module {\sc corner}
    \citep{corner}. }
  \label{fig:corner}
\end{figure*}

The radiative transfer (RT) modeling procedure described in
\citet{Cieza2018} and \citet{Hales2018} is used to derive disk
structural parameters. The model consists of a passively heated disk
characterized by a power-law surface density profile with slope
$\gamma$ and a {\em characteristic radius} $R_{\rm c}$, defined by:
\begin{equation}
  \Sigma(R) = \Sigma_c  \left( \frac{R}{R_c} \right)^{-\gamma} \exp \left[ - \left( \frac{R}{R_c} \right)^{2-\gamma}  \right].
  \label{eq:1}
\end{equation}
The scale height as a function of radius is given by:
\begin{equation}
h(R) = h_c  \left( \frac{R}{R_c} \right)^\Psi ,
\end{equation}
\noindent where $h_c$ is the scale height at the characteristic radius
$R_c$, and $\Psi$ defines the degree of flaring in the disk. Here we
use $H_{100} \equiv h(100~{\rm au})$ to enable comparison with
previous modeling of FUor sources \citep[e.g.,][]{Cieza2018}.
Equation~\ref{eq:1} can be integrated to calculate the disk mass as:
\begin{equation}
M_{\rm d} =  \frac{2 \pi R_c^2  \Sigma_c }{2 - \gamma}.
\end{equation}

This simple disk can therefore be described by 5 free parameters
$M_{\rm d}$, $\gamma$, $R_c$, $H_{100}$ and $\Psi$ (flaring). The flux
emerging from the parametric disk model is computed using the
radiative transfer code {\sc radmc-3d} \citep{Dullemond2012}. Since
FU~Ori is a variable embedded object, the information about its
stellar spectrum is not well known.  Here we assume an effective
stellar temperature of $10^4$~K and a stellar radius of 5~$R_\odot$ to
account for the stellar photosphere and the accretion luminosity. The
computational grid extends from the radius at which the dust
temperature is higher than the sublimation temperature of 1200\,K, up
to an outer radius of 100~au. Since the model is axisymmetric, the
computation is only done in radius and colatitude. The latter extends
from 0 (pole) to $\pi/2$ (midplane) radians. Radial and colatitude
domains are sampled with a grid of 256 by 64 cells, respectively.

We adopt a distribution of dust grains with a power-law in size $a$,
given by $n(a) \propto a^{-3.5}$, extending from 0.1~$\mu$m to
3~mm. For the dust optical properties, we use a mix of amorphous
carbon grains and astrosilicate grains \citep[see][for
  details]{Cieza2018}. The absorption opacity at 1.3 mm is thus
$\kappa_{\rm abs} = 2.2$~cm$^2$g$^{-1}$. The temperature of the dust
particles is calculated using {\sc radmc3d}'s {\tt mctherm}
module. Since we aim to explore thousands of different models, the
modified random walk option is enabled to speed up calculation over
optically thick regions.

The model parameters \{M$_{\rm dust}$, $\gamma$, $R_c$, $H_{100}$,
$\Psi$\} were constrained using a Bayesian approach. In addition to
the disk structure parameters, a centroid shift ($\delta x$, $\delta
y$) is also optimized. The inclination angle $i$ and PA of the model
are fixed to the values obtained from {\tt imfit} (see
Sec.~\ref{sec:image}). This is a compromise we think necessary as the
emission subtends only 2-3 resolution beams. As will be shown in
Sec.~\ref{sec:kine}, this disk orientation is consistent with
signatures of Keplerian rotation seen in $^{12}$CO channel maps.

A 1.3~mm image is produced via ray-tracing, using {\sc radmc3d}'s
second-order integration of the radiative transfer equation. The model
visibilities are interpolated to the same \textit{uv} points as the
observations via a Fast Fourier Transform \citep[we use the same
  algorithm used and described in][]{Cieza2018, Marino2017}. These
model visibilities are compared against the observed visibilities of
each component in FU~Ori.

The posterior distributions of each parameter are sampled with the
{\sc emcee} MCMC algorithm \citep{emcee}. This allows to determine the
set of parameters which maximises the likelihood function (via a
$\chi^2$ comparison computed on the visibility plane). Since the disk
model neglects viscous heating, the derived temperatures serve only as
a crude approximation. Nevertheless, it allows for an estimate of the
size and bulk mass of the dust disk to be obtained.

\subsection{Modeling results}
\label{sec:modelresults}

The results of the MCMC radiative transfer modeling, after running
\revision{2000 iterations} ($\sim$10 times the autocorrelation time)
with 240 walkers, are shown in Fig.~\ref{fig:corner}. The posterior
distributions of the disk structural parameters are single peaked and
relatively narrow, for both components. The 1.3~mm observations of
FU~Ori~N and FU~Ori~S can be described by disk profiles, with
characteristic radii of approximately 11~au for both disks. The slope
of the surface density distribution is $\sim$0.5, also similar for
both disks. The northern component requires a slightly hotter disk
with a scale-height $H_{100}\approx2.8$, compared to the southern dust
emission which requires a scale-height of $\sim$2.3 au at 100 au. See
corner plots in Fig.~\ref{fig:corner} for parameter uncertainties.

\revision{The masses of the dust disks derived from our best radiative
  transfer models are relatively small, with 22$\pm$2~$M_\oplus$ for
  FU Ori~N and 8.8$\pm$1.4~$M_\oplus$ for FU Ori~S.} Our RT
calculations show that the disks around northern and southern
components become optically thick at $r \leq 11$~au and $r \leq 6$~au,
respectively, therefore the inferred total dust masses are likely
underestimated.

This can be explained by the temperatures in our RT calculation, which
reach higher values than the 20~K assumed in previous works. If the
bulk of mass comes from a region at $\sim$50-80~K, the mass estimates
agree within uncertainties.

\section{$^{12}{\rm CO}$ kinematics}
\label{sec:kine}

\begin{figure*}
  \centering\includegraphics[width=\textwidth]{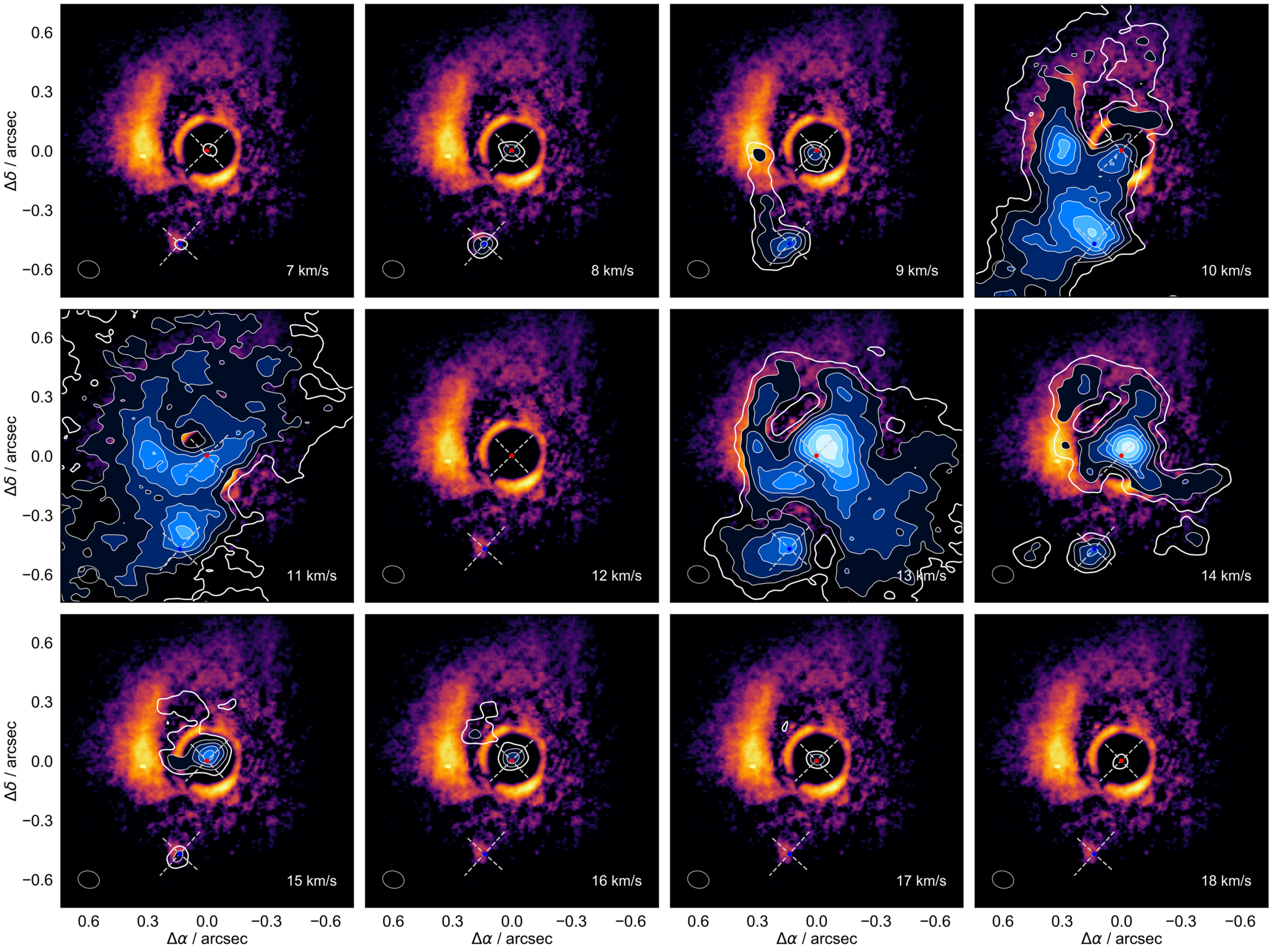}
  \caption{\revision{Complex kinematic environment and evidence of
      disk rotation seen in the $^{12}$CO channel maps of the FU~Ori
      system. The orientation (PA and inclination) inferred from the
      continuum maps are indicated by the white dashed lines for each
      disk. The northern component shows signatures of rotation as
      blue- and red-shifted $^{12}$CO emission loci along the disk
      PA. The red and blue points show the stellar locations of FU Ori
      north and south, respectively. The velocities, in km~s$^{-1}$,
      are given in the lower right corners. Each panel shows $^{12}$CO
      emission via color-filled contour maps for 13, 17, 21, 25, 29,
      33, 37, and 41 mJy~beam$^{-1}$.  The thicker unfilled white
      contour shows the 5$\sigma$ level, i.e., 10 mJy~beam$^{-1}$
      ($\sigma$ is 2 mJy~beam$^{-1}$). The beam size is shown as a
      grey ellipse in the lower left corner. The systemic velocity
      ($\sim$11.7~km~s$^{-1}$) channel at 12~km~s$^{-1}$ suffers heavy
      foreground absorption and displays no emission above
      5$\sigma$. Relative RA and DEC are shown with respect to the
      location of FU~Ori north (peak of the continuum). The background
      image shows the HiCIAO polarized scattered light map published
      in \citet{Takami2018}, see also \citet{Liu2016}.}}
  \label{fig:kine}
\end{figure*}

\subsection{Evidence of disk rotation around each component}

Fig.~\ref{fig:kine} shows the $^{12}$CO channel maps spanning
\revision{11~km~s$^{-1}$} in velocity, centered near FU~Ori~N's
systemic speed of 11.4~km~s$^{-1}$\footnote{The systemic velocity of
  each component was determined via a Gaussian fit to the wings of the
  line profiles integrated over 0\farcs15 apertures centered on each
  star. The systemic speeds of the northern and southern components
  are thus 11.4 and 10.1~km~s$^{-1}$, respectively (with uncertainties
  of about a fifth of the channel width, i.e., 0.2~km~s$^{-1}$.}. The
closest channel to systemic velocity (12 km~s$^{-1}$) suffers from
foreground absorption. The location of each disk is indicated with
white (dashed) crosses, with major axes parallel to the disk PA
inferred from the continuum data.

The local kinematics around the location of FU Ori north shows the
signature of a rotating disk. The emission changes from blue-shifted
(velocities $\leq$11~km~s$^{-1}$, in direction south-east) to
red-shifted ($\geq$13~km~s$^{-1}$, towards north-west). The red- and
blue-shifted emission loci are symmetric with respect to the peak in
dust continuum, along a disk PA of 135$^\circ$ east of north, i.e.,
the disk orientation derived from the continuum emission.

Evidence of spatially-resolved disk rotation around FU~Ori south is
also present in the channel maps, although less clear than the
emission patterns in the vicinity of the northern component. There is
a switch from blue- to red-shifted emission along the southern
continuum disk PA. The switch happens at a systemic speed (for the
southern disk) between the 9 and 10~km~s$^{-1}$ channels. The southern
disk emission is much fainter than its northern counterpart in
channels with velocities $>$11~km~s$^{-1}$. This could be due to
intra-cloud absorption around the southern component. The larger scale
kinematics is complex and heavily influenced by cloud emission and
absorption, and also by a slow outflow (Hales et al. {\it in prep}).

CO channels show a prominent elongated emission feature between 9 and
14~km~s$^{-1}$, immediately east to the primary. At 9~km~s$^{-1}$, the
extended emission is connected (via the lower level flux countour) to
the southern component's kinematics. Yet, at 10 and 11~km~s$^{-1}$ the
emission appears connected to the northern component's blue-shifted
disk kinematics. Interestingly, the brightest feature seen in
reflected light also appears as an arc-like elongation to the east of
FU Ori north \citep{Takami2018}. As shown in Fig.~\ref{fig:kine}, the
extended arc in CO emission indeed could be counterpart to the bright
arc in scattered light. This would imply that the arc-like structure
could be moving at (projected) speeds of $\sim$2-3~km~s$^{-1}$, with
respect to the north component's systemic speed.

Although using wide spectral channels allows to pinpoint disk
kinematics, this resolution is not ideal to determine the kinematics
of the bright blue arc. To study the nature of this feature, whether
it is an outflowing or inflowing structure, requires deep gas
observations in optically thinner tracers.
 
\begin{figure}
  \centering\includegraphics[width=.8\columnwidth]{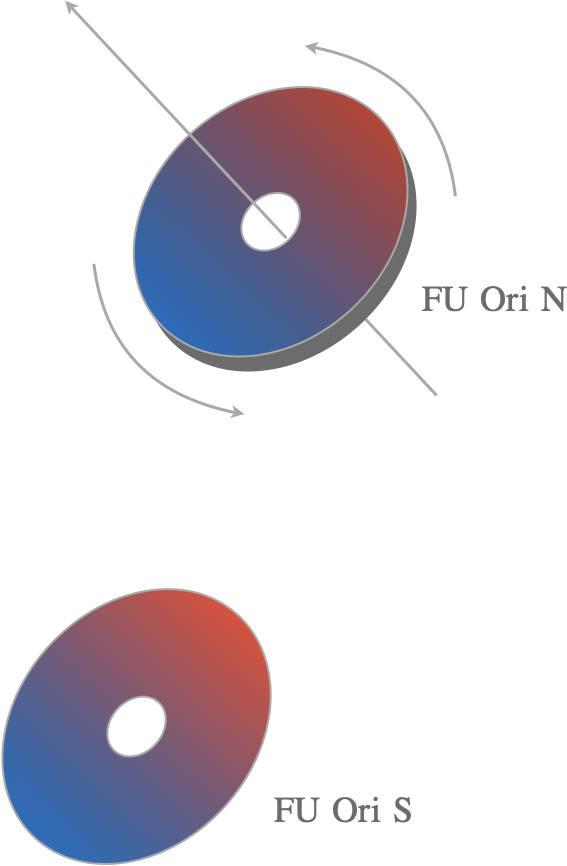}
  \caption{Cartoon representation showing the geometry and the
    kinematics of the FU Ori system.  The disks are scaled up to aid
    visualization. The inclination and position angles are as inferred
    from the continuum image. The kinematics (blue/red gradient) is
    consistent with the $^{12}$CO channel maps. The inner holes are
    drawn for aesthetic reasons.}
  \label{fig:cartoon}
\end{figure}

\revision{The presence of a blue-shifted large scale outflow to the
  north-east of FU Ori, as well as a red-shifted counterpart to the
  south-west, will be presented in a future work (Hales et al. {\em in
    prep}). Assuming that the outflow is launched from the FU~Ori~N
  disk implies that the near side of the disk is to the
  south-west. The evidence for rotation seen in the $^{12}$CO channel
  maps allows to infer counter-clockwise disk rotation in the plane of
  the sky. Figure \ref{fig:cartoon} shows a cartoon representation of
  the inferred geometries of the FU~Ori binary. The inclinations and
  position angles correspond to those calculated from the continuum
  image. The kinematics of each disk, inferred from the $^{12}$CO
  channel maps, is shown as a red/blue gradient. In the case of
  FU~Ori~S, the disk's near/far side and the sense of rotation are
  unknown.}

\section{Discussion}
\label{sec:results}

\subsection{FU Ori disk sizes and masses}

Modeling of the continuum emission yields disk sizes that are
$\sim$11~{\sc au} in characteristic radius for both components. These
small disk sizes are comparable to those of protostellar disks, which
have typical radii of less than 10~au at 8~mm \citep{SeguraCox2018}.
The slope of the density distribution of each FU~Ori disk is
$\sim$0.4, closer to TTauri disks \citep{Andrews2010} than other FUor
sources observed at 1.3 mm \citep{Cieza2018}. The dust masses and disk
sizes inferred from the RT modeling are small compared to the
continuum emission from other FUor and EXor sources. \citet{Cieza2018}
fit the continuum emission of eight FUor and EXor targets (their
figure 6), and found that FUor are more massive than EXor.
Interestingly, our resolved observations of FU~Ori north and south
suggest that the FUor prototype system have dust masses comparable to
the EXor sources in \citet{Cieza2018}'s sample.


\subsection{Update on the FU Ori accretion disk model}

With the assumption that the inclination of FU~Ori~N is 55$^\circ$,
previous SED fitting to its optical-IR spectrum suggests that FU~Ori~N
harbors a disk accreting at a rate of $\dot{M}\approx2.4\times
10^{-4}$ M$_{\odot}$yr$^{-1}$ around a 0.3 M$_{\odot}$ star
\citep{Zhu2007}. The inner radius is 5~R$_{\odot}$ in this accretion
disk model. Our observations presented here suggest an inclination
angle of $\sim$35$^\circ$. This new inclination allows us to update
the disk and star parameters for FU~Ori~N. First, to produce the same
line-width that explains the optical spectrum of FU Ori
\citep{Zhu2009a}, the stellar mass has to increase to
0.6~M$_{\odot}$. With the smaller inclination angle, the disk
luminosity is only 70\% of the previous estimate, which leads to 84\%
of the previous disk inner radius and 59\% of the previous
$M\,\dot{M}$ value. \revision{Thus, the disk inner radius is
  3.5~R$_{\odot}$ and $\dot{M}$ is 3.8$\times10^{-5}$ M$_{\odot}$
  yr$^{-1}$ with the new central star mass.}

Given the very low mass of the disks and the high accretion rate
inferred above, the accretion event must last for a short time
compared to the disk lifetime. The high accretion rate suggests that
the mechanism behind the outbursts in luminosity happens in an
episodic way. \revision{There is the possibility that the disk mass is
  replenished efficiently by inflow of material from the cloud or by
  cloudlet capture \citep[e.g.,][]{Dullemond2019}. Our observations do
  not rule out the presence of inflowing material and we hope to
  further explore this scenario in follow-up observations connecting
  the large scale outflows with higher sensitivity and resolution
  local kinematics}.  In the following section we discuss the
possibility of interaction scenarios.

\subsection{Dynamically perturbed kinematics?}

The disk rotation pattern around both components are skewed, in the
sense that a unique PA is not sufficient to represent the emission
loci over the full velocity range. In the FU~Ori binary system, this
could suggest that the disks are being perturbed by mutual
interactions. \revision{An internal perturbation, e.g. driven by
  self-gravity, is unlikely given the low mass of the disks. Given the
  proximity of both stars, we think that gravitational interactions
  are likely to be at play.} Although the dust disks share similar
geometries, the orbit of the perturber, FU Ori south, need not be in
the same plane. The bright arc, seen in scattered light and CO in
Fig.~\ref{fig:kine} may be explained as a dynamical feature out of the
plane of the FU Ori north disk if created by flyby or disk-disk
encounter \citep[see][, their fig. 4]{2019MNRAS.tmp.2545C}.

The compact size of the dust disks can also be explained within a
flyby scenario. An (inclined) prograde encounter, besides stripping
the disk of dust material in the outer regions, leads to an
enhancement of the density in the inner regions of the disk \citep[see
  red curves in][their figure 11]{Cuello2019}. This prograde encounter
could also potentially explain the outburst event via enhanced
accretion. However, in this case the perturbation only happens once.

\revision{An alternative scenario to a disk-disk interaction has been
  proposed by \citet{Dullemond2019}. Here, the capture of a cloudlet
  or cloud fragment also leads to arc-shaped reflection nebulae. The
  capture of this cloud fragment also replenishes the disk allowing
  for a fresh supply of material to maintain the high accretion rate.}


\section{Concluding remarks}
\label{sec:conclusions}

The new 1.3 mm ALMA data presented here resolve the continuum and gas
emission around both FU~Ori components. We have derived sizes and
orientations for each disk from continuum emission. The disk
orientations suggest that FU Ori north and south share similar
inclination angles. If we were to assume the disk are in near coplanar
configuration, they are separated by a deprojected distance of 250~au.
Their sizes are remarkably similar taking into account that the
southern stellar component is at least twice as massive as the
northern star \citep{beck2012}.

The spatially-resolved $^{12}$CO kinematics allowed for disk rotation
to be identified in the vicinity of each component. The emission
revealing disk rotation also appears asymmetric and skewed, suggesting
the disks are subject to interaction \revision{in the form of a
  flyby. This interaction could be due to a mutual disk-disk encounter
  and/or cloudlet capture within the cloud. Moreover, the slow
  channels (near systemic velocity) reveal CO emission which is
  spatially coincident with the bright reflection arc previously
  reported in scattered light observations. The elongated feature is,
  to some extent, connected to both the north and south components. To
  understand whether this emission corresponds to inflow or outflow to
  or from any of the FU Ori components will require deeper CO
  observations to sample the kinematics with finer spectral resolution
  and higher sensitivity.} The gas observations also show that the
systemic speeds of FU Ori north and south differ by at least
1~km~s$^{-1}$, which could inform future dynamical models of the
binary interaction. Several FUor type objects have known binary
companions, thus investigating the kinematics around these multiple
systems can help connect binary interaction and flybies with accretion
bursts triggers.

The disks in FU Ori north and south are compact and mostly optically
thick, even at mm wavelengths \citep[similarly to Class
  I,][]{sheehan2017, SeguraCox2018}.  In order to determine the disk
masses and dust properties within these central regions, observations
at longer wavelengths with {\sc au}-scale resolutions will be required
\citep[i.e., at next-generation Very Large Array
  baselines,][]{white2018, murphy2018}.

\acknowledgments

We thank the anonymous referee for a constructive report. We also
thank Sebasti\'an Marino for providing the tool to sample the model in
visibilities, Ed Fomalont for useful discussions and suggestions
during the analysis of the data, and Michihiro Takami and Jun
Hashimoto for providing the HiCIAO image, as well as useful
comments. SP acknowledges support from CONICYT-Gemini grant 32130007,
CONICYT-Fondecyt Regular grants 1191934, and the Joint Committee of
ESO and the Government of Chile. H.B.L. is supported by the Ministry
of Science and Technology (MoST) of Taiwan (Grant
Nos. 108-2112-M-001-002-MY3 and 108-2923-M-001-006-MY3). NC
acknowledges financial support provided by FONDECYT grant 3170680. We
acknowledge support from the Millennium Science Initiative (Chile)
through grant RC130007. This work used the Brelka cluster, financed by
Fondequip project EQM140101, hosted at DAS/U. de Chile.  This paper
makes use of the following ALMA data: {\sc
  ADS/JAO.ALMA.2016.1.01228.S}. ALMA is a partnership of ESO
(representing its member states), NSF (USA) and NINS (Japan), together
with NRC (Canada) and NSC and ASIAA (Taiwan), in cooperation with the
Republic of Chile. The Joint ALMA Observatory is operated by ESO,
AUI/NRAO and NAOJ. The National Radio Astronomy Observatory is a
facility of the National Science Foundation operated under cooperative
agreement by Associated Universities, Inc.

\facility{ALMA}.

\software{emcee \citep{emcee}, radmc-3d \citep{Dullemond2012}, Common
  Astronomy Software Applications \citep{2007ASPC..376..127M}, astropy
  \citep{2013A&A...558A..33A}}

\end{document}